\documentstyle[emulateapj,apjfonts]{article}

\newcommand{\Ly}{Ly$\alpha$ }

\newcommand{\ngc}{NGC~1068 }
\newcommand{\ngcs}{NGC~1068's }
\newcommand{\ngcb}{NGC~1068}
\newcommand{\ha}{$\rm{H}\alpha$ }
\newcommand{\ebv}{$\rm{E(B - V)}$ }
\newcommand{\rv}{$\rm{R_{v}}$ }

\lefthead{GRIMES, KRISS, \& ESPEY}
\righthead{SPATIALLY RESOLVED HUT SPECTRA OF NGC 1068}

\begin{document}
\submitted{Accepted for publication in the Astrophysical Journal 6 July 1999}

\title{Spatially Resolved Hopkins Ultraviolet Telescope Spectra of NGC~1068}

\author{John P. Grimes\altaffilmark{1,2},
Gerard A. Kriss\altaffilmark{1,3},
Brian R. Espey\altaffilmark{3}}

\authoremail{jpgrimes@midway.uchicago.edu, gak@stsci.edu, espey@stsci.edu}

\altaffiltext{1}{Department of Physics and Astronomy, Johns Hopkins University,
Baltimore, MD 21218-2695}
\altaffiltext{2}{Department of Physics, University of Chicago,
Chicago, IL 60637}
\altaffiltext{3}{Space Telescope Science Institute, 3700 San Martin Drive,
Baltimore, MD 21218}

\begin{abstract}
We present spatially resolved far-ultraviolet spectra (912--1840~\AA) of
NGC~1068 obtained using the Hopkins Ultraviolet Telescope (HUT) 
during the March 1995 Astro-2 mission.  
Three spectra of this prototypical Seyfert 2 galaxy were obtained 
through a 12\arcsec~diameter aperture centered on 
different locations near the nucleus.  The first pointing 
(A1) was centered west of the optical nucleus; the nucleus was on
the eastern edge of the aperture.  The second (A2) was centered 
southwest of the optical nucleus with the nucleus well inside the aperture.
The third (B) was centered on the ionization cone, with the nucleus 
on the southwestern edge of the aperture.  
While all three aperture locations have spectra similar to the Astro-1
observations of Kriss et al., these new spatially resolved observations
localize the source of the far-UV line and continuum emission.

The ionization cone (location B) has both brighter emission lines and
continuum than the nuclear region (location A2).
Position A1 is fainter than either A2 or B in both lines and continuum.  
The far-UV emission lines observed with HUT have a spatial distribution that
most closely resembles that of [\ion{O}{3}] $\lambda$5007, but appear to be
more extended and offset to the northeast along the axis of the radio jet.
This supports the previous conclusion of Kriss et al. that the
bright {\sc C~iii} $\lambda 977$ and {\sc N~iii} $\lambda 991$
arises in shock-heated gas.

The UV continuum radiation has a more extended spatial distribution than
the line-emitting gas.  At wavelengths longward of 1200 \AA\ the inferred
continuum distribution is consistent with that seen in archival
Hubble Space Telescope WFPC2 images through filter F218W, and it
appears to contain a substantial contribution from starlight.
At wavelengths shorter than 1200 \AA, the UV continuum becomes more
concentrated in a region matching the location and shape of the
UV line radiation, consistent with nuclear flux scattered by a combination
of the electron scattering mirror and the NE dust cloud.

\end{abstract}


\section{Introduction}

The proximity, brightness, and rich phenomenology of NGC~1068 have made it a
key source for our current understanding of the structure and physics of
active galactic nuclei (AGN).
The bright, narrow, high-excitation emission lines of NGC~1068 define it as
the prototype of the Seyfert 2 class.
In polarized light, however, it shows the blue continuum and broad permitted
emission lines typical of Seyfert 1s (\cite{Antonucci85};
\cite{MGM91}; \cite{Code93}; \cite{Antonucci94}).
These characteristics inspired the ``unified model" of AGN in which the
different types of Seyfert galaxy result from a combination of orientation,
obscuration, and reflection of light from the continuum source
and the broad-line region (BLR). (See the review by \cite{Antonucci93}.)
For Seyfert 2 galaxies, the observer's line of sight lies near the plane of
an opaque torus that blocks a direct view of the continuum source and the BLR.
Electrons in clouds of hot gas and dust in cooler clouds above and below the
plane of the torus reflect radiation from central regions into the
observer's line of sight.
For Seyfert 1 galaxies, the observer's line of sight lies well above the plane
of the torus, resulting in an unobstructed view of the interior.

\begin{figure*}[b]
\plotfiddle{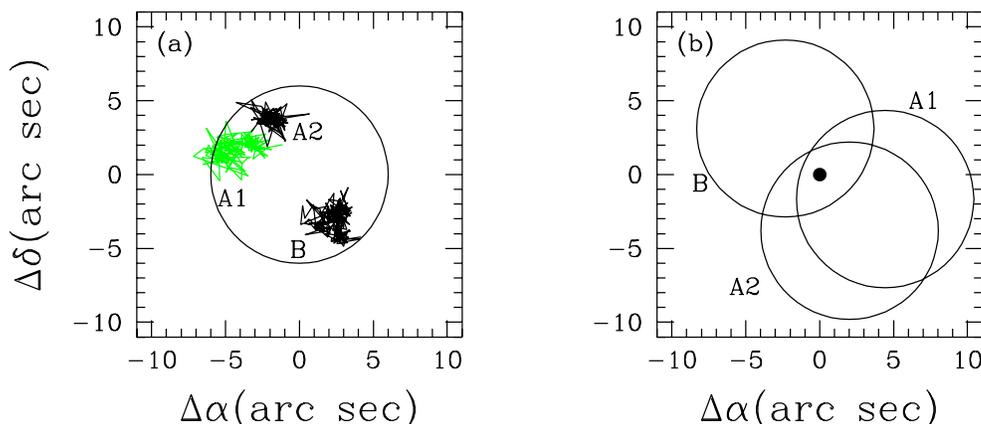}{2.0 in} {0}{70}{70}{-225}{-290}
\caption{
\footnotesize
HUT pointing errors during the observations of \ngc are obtained from
guide stars in the field of the acquisition TV camera.
In the left panel (a) we plot the location of \ngcs optical nucleus relative
to the center of the HUT 12\arcsec\ aperture.
The labels A1, A2, and B identify the data groupings used for the
spectra discussed in this paper.
The right panel (b) shows the locations determined from the centroids of the
pointing errors for aperture positions A1, A2, and B relative to the optical
nucleus.
\label{n1068f1.ps}}
\end{figure*}

The obscuring torus not only blocks radiation from reaching an observer, but it
also shadows gas in the surrounding regions of the galaxy.
This anisotropic illumination can produce conical emission-line regions
frequently referred to as ``ionization cones" (\cite{Pogge89};
\cite{Tsvetanov89}; \cite{Evans91}, 1993\markcite{Evans93},
1994\markcite{Evans94}).
This standard interpretation presumes that photoionization by radiation
from the central source is the primary energy input into the
narrow-line region (NLR).
The observed line ratios corroborate this interpretation when compared to
photoionization models (e.g., \cite{FO86}; \cite{VO87}; \cite{BCR88}).
However, in addition to radiation, kinetic energy in the form of outflowing
winds and radio jets may also play a significant role in transferring energy
from the nuclear region into the surrounding galaxy along the axis of the torus.
The principal reflecting region for Seyfert 2s is most likely a wind of hot
($10^5$ K) electrons driven off the torus by X-rays from the central source
(\cite{KB86}; \cite{KL89}).
Radio jets in Seyferts are also preferentially aligned with the axis of the
ionization cones (\cite{WT94}).
Kinetic energy from these sources may be a significant input to the energy
budget of the NLR.  A number of authors have suggested that shocks from such
interactions may power a large fraction of the line emission.
Morse, Raymond, \& Wilson (1996)\markcite{MRW96}
review the status of shocks for ionizing gas in the NLR.
In cases like the bow shock models of
Wilson \& Ulvestad (1987)\markcite{Wilson87}
and Taylor, Dyson, \& Axon (1992)\markcite{Taylor92},
shocks compress the gas and enhance its radiative output, but
nuclear radiation drives the ionization.
In the ``autoionizing shock" models of Sutherland, Bicknell, \&
Dopita (1993)\markcite{Sutherland93} and
Dopita \& Sutherland (1995)\markcite{Dopita95},
ionizing photons generated in the primary
shocks themselves photoionize the surrounding gas.

A key observational feature of the autoionizing shock models is the strength
of collisionally excited far-UV emission lines.  Lines such as {\sc O~vi}
$\lambda\lambda 1032,1037$, {\sc C~iii} $\lambda 977$, and {\sc N~iii}
$\lambda 991$ have high excitation temperatures and are thus prime
coolants in the high temperature regions of fast shocks.
These lines are particularly strong in NGC~1068 as seen in HUT observations
during the Astro-1 mission (Kriss et al. 1992), and the temperature-sensitive
ratios I({\sc C~iii}] $\lambda 1909$)/I({\sc C~iii} $\lambda 977$) and
I({\sc N~iii}] $\lambda 1750$)/I({\sc N~iii} $\lambda 991$)
implied temperatures exceeding 50,000~K--- temperatures far higher than
those characteristic of thermally stable photoionized gas.
However, Ferguson, Ferland, \& Pradhan (1995)\markcite{FFP95} argued that
strong {\sc C~iii} $\lambda 977$ and {\sc N~iii} $\lambda 991$
could arise from fluorescence in photoionized gas if turbulent velocities
exceeded $\sim$1000 $\rm km~s^{-1}$.

Radio structures in NGC~1068 show a strong spatial correlation with
the emission line gas at visible wavelengths, e.g.
[{\sc O~iii}] $\lambda 5007$ (\cite{Wilson87};
\cite{Evans91}; \cite{Gallimore96}; \cite{Capetti97}), and in the
near-infrared, e.g., [Fe~{\sc ii}] 1.6435 $\mu$m (\cite{Blietz94}).
The Astro-1 HUT spectra of NGC~1068 lacked spatial resolution on scales
smaller than the $18''$ and $30''$ circular apertures used for the observations.
Neff et al. (1994\markcite{Neff94}) presented far-UV images with
$\sim 2$\arcsec\ resolution obtained with the Ultraviolet Imaging Telescope
(UIT) on the Astro-1 mission, but this broad band (1250--2000 \AA) image
did not separate line and continuum emission.
To obtain information on the spatial distribution
of the far-UV emission lines and continuum flux, we carried out
the observations described in this paper during the Astro-2 mission.
We compare our spatially resolved spectra to the far-UV UIT images and
to emission line and continuum images
at longer wavelengths obtained with HST (\cite{Dressel97}).
We find that the far-UV emission lines observed with HUT
are more extended than the [{\sc O~iii}] $\lambda 5007$ emission observed
with HST, and that it is offset to the northeast along the direction
of the radio jet.
At wavelengths greater than 1200 \AA, the UV continuum has a greater spatial
extent than the emission lines.  At shorter wavelengths, it becomes more
spatially concentrated.

In sections 2 and 3 we describe the HUT Astro-2 observations and our data
reduction process.
We then discuss in \S4 the spatial information that can be gleaned from the
HUT observations.
In \S5 we compare the HUT observations to the UIT and HST images.
We discuss the implications of our observations for the excitation of the
line emission in NGC~1068 and for the origin of the ultraviolet continuum
in \S6.  We summarize our conclusions in \S7.

\section{HUT Observations}

During the course of the 16-day Astro-2 space shuttle mission in 1995 March
we used HUT to obtain one-dimensional spectra through a 12\arcsec\ aperture at
three distinct spatial locations in the nuclear region of NGC~1068.
HUT uses a 0.9-m primary mirror in conjunction with a prime-focus,
Rowland-circle spectrograph to obtain spectra with a resolution of
$\sim$3 \AA\ spanning the 820--1840 \AA\ band.
The primary mirror and the concave grating are both coated with SiC to
provide high UV reflectivity at wavelengths shortward of 1200 \AA.
Light dispersed by the prime-focus grating is focussed onto a photon-counting
detector consisting of a micro-channel-plate intensifier with a CsI
photocathode and a phosphor-screen anode.
A 1024-diode linear Reticon array is used to detect the intensified pulses on
the anode.
Events are centroided to a half-diode precision, producing a 2048-pixel,
one-dimensional spectrogram.
Davidsen et al. (\markcite{Davidsen92}1992)
provide a detailed description of HUT.
Improvements made to HUT for the Astro-2 mission and HUT's
in-flight performance are described by Kruk et al. (\markcite{Kruk95}1995).

\begin{deluxetable}{lcccl}
\tablecolumns{5}
\tablewidth{0pc}
\tablecaption{HUT Astro-2 Observations of \ngc
\label{tbl:observation} }
\tablehead{
\colhead{Observation} & \colhead{UT Date} & \colhead{UT Start} &
\colhead{Integration Time} & \colhead{Comments} \\
\colhead{} & \colhead{} & \colhead{} & \colhead{(s)} & \colhead{}}
\startdata
A1d & 5 March 1995 & 02:43:28 & 716 & Day   \\
A1n & 5 March 1995 & 02:55:29 & 290 & Night \\
A2  & 5 March 1995 & 03:00:32 & 660 & Night \\
Bd  & 7 March 1995 & 15:47:53 & 496 & Day   \\
Bn  & 7 March 1995 & 15:56:14 & 888 & Night \\
\enddata
\end{deluxetable}

The HUT guidance system relies on a slit-viewing video camera and guide stars.
Using this information HUT is manually pointed by a
payload specialist aboard the space shuttle.
Video frames taken during the observation enable later reconstruction
of HUT's pointing during the observation.
Positions of guide stars in the video frames can be centroided to an accuracy
of $\sim 0.5''$ (about 0.5 pixels).  The aperture position in the field of
view also changes slightly between observations as the slit wheel rotates.
Its position can be measured to an accuracy of $\sim 0.5 ''$ by measuring the
apparent hole it leaves in images with a bright background, such as those
obtained near the earth limb in orbital daylight.
In Figure~\ref{n1068f1.ps} we show the pointing errors derived from the video
images during our observations of NGC~1068.
These errors give the location of the optical nucleus relative to the
center of the HUT $12''$ aperture.

Three dominant groupings of pointing errors are apparent in
Figure~\ref{n1068f1.ps}, and we use these as the basis for the three separate
spectra we discuss here.
During the course of the first observation (A), 
the payload specialist moved the aperture $\sim3\arcsec$ southwest
to place the optical nucleus definitely within the slit.
As this was a fairly substantial pointing correction, we have split this
observation into two separate pieces.
During A1 the optical nucleus is on the very edge of the 
slit and in A2 the optical nucleus is well within the slit.
The second observation (B) occurred one day later.
The ionization cone was centered within the aperture while
the optical nucleus was near the southwest edge of the aperture.

\section{HUT Data}

\begin{figure*}[t]
\plotfiddle{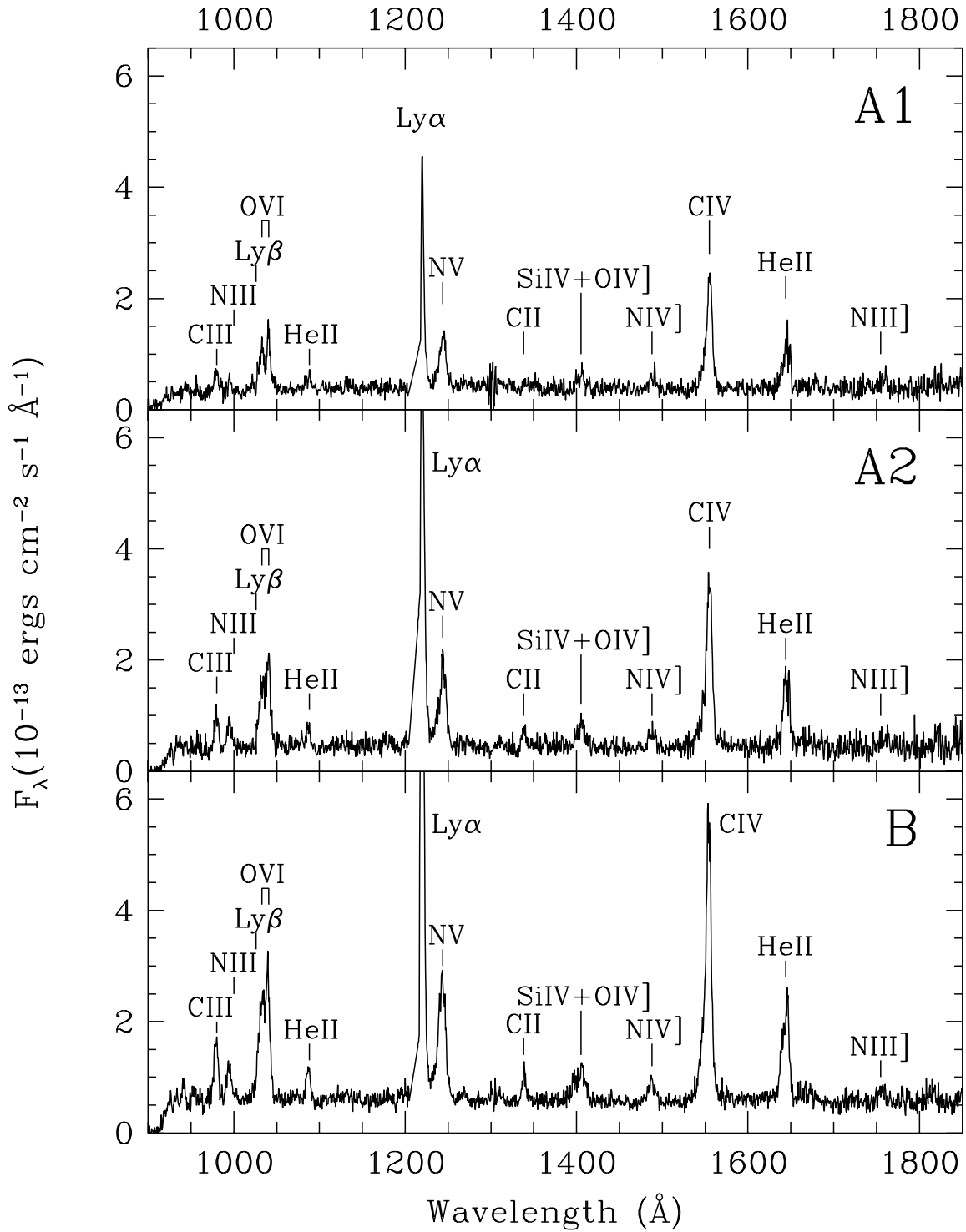}{7.5 in} {0}{85}{85}{-260}{-76}
\caption{
Flux-calibrated HUT spectra from the Astro-2 observations.
Observation A1 (top panel) had the optical nucleus on the eastern edge of
a 12$''$-diameter aperture.
Observation A2 (middle panel) placed the optical nucleus inside the aperture
with the ionization cone region and radio jet outside the northeast edge.
Observation B (bottom panel) was centered on the ionization
cone.  The optical nucleus was on the southwest edge of the aperture.
\label{n1068f2.ps}}
\end{figure*}

Due to the differences between night 
and day airglow we separated the night and day portions
of observations A1 and B for independent processing.
All of observation A2 occurred during orbital night.
These data groupings are summarized in Table \ref{tbl:observation}. 
We reduced the data using the standard procedures described
by Kruk et al. (\markcite{Kruk95}1995; \markcite{Kruk99}1999).
We determined background from regions free of airglow at wavelengths
shortward of the 912 \AA\ Lyman limit.
With the exception of the extremely strong geocoronal \Ly line at 1216 \AA\ we
fitted the airglow lines with symmetric Gaussians and then subtracted the fitted
profiles.
Fitting the geocoronal \Ly line is more difficult due to its
broad scattering wings.  Thus we constructed \Ly
templates from blank field observations taken during the Astro-2 mission.
We then subtracted the appropriately scaled \Ly profile, including
its broad scattering wings, from each spectrum.
We flux calibrated the background-subtracted spectra by applying
a time-dependent inverse-sensitivity curve derived from HUT observations and
atmospheric models of white dwarfs (\cite{Kruk95}; \cite{Kruk99}).
A minor correction for second-order light in the 1824-1840 \AA~range was
made based on the measured intensity of the 912--920 \AA\ region.
The errors for the raw count spectra were calculated by
assuming a Poisson distribution.
These errors were propagated through the reduction process.
Figure \ref{n1068f2.ps}
shows our final calibrated spectra for each of the HUT aperture locations.

\setcounter{footnote}{0}
We use the IRAF\footnote{
The Image Reduction and Analysis Facility (IRAF)
is distributed by the National Optical Astronomy
Observatories, which is operated by the Association of
Universities for Research in Astronomy,Inc. (AURA),
under cooperative agreement with the National 
Science Foundation.}
task {\tt specfit} (\cite{Kriss94}) to fit the 
continuum and emission lines in our spectra.
For the continuum we assume a power law shape of the form 
$\rm{F}_{\lambda}=\rm{F}_{0} (\lambda/1000)^{-\alpha}$
modified by the extinction curve of Cardelli, 
Clayton, \& Mathis (1989\markcite{Cardelli89}).
Keeping \rv fixed at 3.1, we found \ebv= 0.02
provides the best fit for all three spectra.  The best fit
power law normalization and spectral indices are listed in
Table 2.  Our best fit extinction is in 
rough agreement with those used in previous 
ultraviolet studies of \ngc (\cite{Kriss92}, 
\cite{Snijders86}, and \cite{FFP95}).  
Fortunately, due to the comparative nature of our study, our
results are insensitive to uncertainties in the extinction.

\begin{center}
\vbox to 10pt {\vfill}
{\sc Table 2\\
Continuum Parameters for the HUT Spectra of NGC~1068$\rm ^a$}
\vskip 4pt
\begin{tabular}{lcc}
\hline
\hline
Observation & $\rm{F}_{0}$ & $\alpha$ \\
\hline
A1 & 4.939 & 0.240 \\
A2 & 5.834 & 0.294 \\
B  & 8.070 & 0.415 \\
\hline
\end{tabular}
\vskip 2pt
\parbox{3.5in}{
\footnotesize
$\rm ^a$The parameters
describe a power law $\rm{F}_{\lambda}=\rm{F}_{0}(\lambda/1000)^{-\alpha}$
with $\rm{F}_{0}$ in units of $\rm{10^{-14}~ergs~cm^{-2}~s^{-1}~\AA^{-1}}$,
assuming an extinction correction of \ebv= 0.02.
}
\vbox to 10pt {\vfill}
\end{center}
\setcounter{table}{2}

To measure the emission lines, we fit symmetric Gaussians to the 
spectral features of \ngcb.
In our fits we have not included the wavelength region from 1200-1217 \AA\ due
to contamination from the \Ly geocoronal line.
However \ngcs intrinsic \Ly emission line was sufficiently bright and redshifted
to disentangle it from geocoronal emission.
We detected and fit broad components to the
\Ly and \ion{C}{4} $\lambda\lambda1548,1551$ emission lines
as described by Kriss et al. (1992)\markcite{Kriss92}.
In our fits we linked the widths of the \Ly and \ion{C}{4} broad line components
so that they were identical.
We also linked the widths and redshifts of the blended
Ly$\beta~\lambda1026$ and \ion{O}{6} $\lambda\lambda1032,1038$ lines.

Tables \ref{tbl:A1}--\ref{tbl:B} give the best fit
fluxes, velocities, and full-width at half-maximum (FWHM)
for the emission lines in observations A1, A2, and B.
The fluxes are as observed, with no correction for extinction.
The velocities are relative to a systemic redshift of 
$z = 0.0038$ (\cite{Huchra92}).
The quoted error bars are the formal 
$1 \sigma$ uncertainties derived from the error matrix of our fit.
We caution that errors for observation A1 could be underestimated due to a high
percentage of day airglow contamination
and relatively large pointing deviations for this observation.
Our flux measurements and Gaussian widths are consistent with earlier data from 
Astro-1 (\cite{Kriss92}) and IUE (\cite{Snijders86}).
The widths of our \Ly and \ion{C}{4} broad lines are in excellent agreement
with the H$\beta$ width measured by
Miller et al. (1991;\markcite{Miller91} FWHM=3030~$\rm km s^{-1}$).
As in the earlier observations by Kriss et al. (\markcite{Kriss92}1992), we were
unable to detect the \ion{O}{3}] $\lambda\lambda 1660,1666$ lines in any of our
observations.

\begin{deluxetable}{lrccc}
\tablecolumns{5}
\tablewidth{0pc}
\tablecaption{Emission Lines in HUT Observation A1 of \ngc
\label{tbl:A1} }
\tablehead{
        \colhead{} & 
        \colhead{$\rm{\lambda_{vac}}$} & 
        \colhead{Flux} & 
        \colhead{$\Delta \upsilon$\tablenotemark{a}} & 
        \colhead{FWHM} \\
        \colhead{Line} & 
        \colhead{$\rm{(\AA)}$} & 
        \colhead{$\rm{(10^{-13}~ergs~cm^{-2}~s^{-1})}$} & 
        \colhead{$\rm{(km~s^{-1})}$} & 
        \colhead{$\rm{(km~s^{-1})}$} }
\startdata
C~III          & \phantom{0}977.03 & $\phantom{-}1.9 \pm~0.3$ & \phantom{0}$-215 \pm \phantom{0}154$ & $\phantom{-}1904 \pm \phantom{0}409$ \nl
N~III          & \phantom{0}991.00 & $\phantom{-}0.9 \pm~0.2$ & $\phantom{-0}228 \pm \phantom{0}102$ & $\phantom{-}1201 \pm \phantom{0}214$ \nl
Ly$\beta$      & 1025.72 & $\phantom{-}2.9 \pm~0.6$ & \phantom{00}$-52 \pm \phantom{0}157$ & $\phantom{-}1782 \pm \phantom{0}302$ \nl
O~VI           & 1031.93 & $\phantom{-}3.1 \pm~0.3$ & \phantom{00}$-52 \pm \phantom{0}157$ & $\phantom{-}1782 \pm \phantom{0}302$ \nl
O~VI           & 1037.62 & $\phantom{-}5.1 \pm~0.4$ & \phantom{00}$-52 \pm \phantom{0}157$ & $\phantom{-}1782 \pm \phantom{0}302$ \nl
He~II          & 1085.15 & $\phantom{-}1.8 \pm~0.3$ & \phantom{0}$-455 \pm \phantom{0}157$ & $\phantom{-}2166 \pm \phantom{0}345$ \nl
Ly~$\alpha$~n  & 1215.67 & $11.5 \pm 2.1$ & $\phantom{-00}15 \pm \phantom{00}31$ & $\phantom{-0}849 \pm \phantom{0}120$ \nl
Ly~$\alpha$~b  & 1215.67 & $10.9 \pm 3.3$ & \phantom{0}$-565 \pm \phantom{0}145$ & $\phantom{-}2843 \pm \phantom{0}239$ \nl
N~V            & 1240.15 & $10.2 \pm~0.4$ & \phantom{0}$-263 \pm \phantom{00}52$ & $\phantom{-}2645 \pm \phantom{0}123$ \nl
C~II           & 1334.53 & $\phantom{0}0.7 \pm~0.2$ & $\phantom{-0}199 \pm \phantom{0}295$ & $\phantom{-}1428 \pm \phantom{0}533$ \nl
Si~IV+O~IV]    & 1400.00 & $\phantom{-}2.5 \pm~0.3$ & $\phantom{-00}13 \pm \phantom{0}141$ & $\phantom{-}2261 \pm \phantom{0}286$ \nl
N~IV]          & 1486.50 & $\phantom{-}2.0 \pm~0.3$ & \phantom{0}$-654 \pm \phantom{0}124$ & $\phantom{-}1791 \pm \phantom{0}275$ \nl
C~IV~n         & 1549.05 & $10.1 \pm 1.8$ & $\phantom{-00}59 \pm \phantom{00}38$ & $\phantom{-}1217 \pm \phantom{0}137$ \nl
C~IV~b         & 1549.05 & $\phantom{-}8.6 \pm 1.8$ & \phantom{0}$-450 \pm \phantom{0}145$ & $\phantom{-}2843 \pm \phantom{0}239$ \nl
He~II          & 1640.50 & $\phantom{-}8.1 \pm~0.5$ & \phantom{0}$-392 \pm \phantom{00}55$ & $\phantom{-}1817 \pm \phantom{0}119$ \nl
O~III]          & 1664.00 & $<1.5 $ & $\phantom{-}1243 \pm \phantom{0}240$ & $\phantom{-}1817 \pm \phantom{0}119$ \nl
N~III]         & 1750.00 & $\phantom{-}1.7 \pm~0.4$ & $\phantom{-0}227 \pm \phantom{0}102$ & $\phantom{-}1201 \pm \phantom{0}214$ \nl
\enddata
\tablenotetext{a}{Velocities are relative to a systemic redshift of 
$z = 0.0038$ (\cite{Huchra92}).}
\end{deluxetable}

\begin{deluxetable}{lrccc}
\tablecolumns{5}
\tablewidth{0pc}
\tablecaption{Emission Lines in HUT Observation A2 of \ngc
\label{tbl:A2} }
\tablehead{
        \colhead{} & 
        \colhead{$\rm{\lambda_{vac}}$} & 
        \colhead{Flux} & 
        \colhead{$\Delta \upsilon$\tablenotemark{a}} & 
        \colhead{FWHM} \\
        \colhead{Line} & 
        \colhead{$\rm{(\AA)}$} & 
        \colhead{$\rm{(10^{-13}~ergs~cm^{-2}~s^{-1})}$} & 
        \colhead{$\rm{(km~s^{-1})}$} & 
        \colhead{$\rm{(km~s^{-1})}$} }
\startdata
C~III          & \phantom{0}977.03 & $\phantom{-}3.2 \pm~0.3$ & \phantom{0}$-260 \pm \phantom{00}86$ & $\phantom{-}1613 \pm \phantom{0}176$ \nl
N~III          & \phantom{0}991.00 & $\phantom{-}3.2 \pm~0.4$ & $\phantom{-0}146 \pm \phantom{0}120$ & $\phantom{-}2212 \pm \phantom{0}246$ \nl
Ly$\beta$      & 1025.72 & $\phantom{-}3.6 \pm~0.4$ & \phantom{0}$-139 \pm \phantom{00}82$ & $\phantom{-}1771 \pm \phantom{0}129$ \nl
O~VI           & 1031.93 & $\phantom{-}6.9 \pm~0.4$ & \phantom{0}$-139 \pm \phantom{00}82$ & $\phantom{-}1771 \pm \phantom{0}129$ \nl
O~VI           & 1037.62 & $\phantom{-}8.5 \pm~0.5$ & \phantom{0}$-139 \pm \phantom{00}82$ & $\phantom{-}1771 \pm \phantom{0}129$ \nl
He~II          & 1085.15 & $\phantom{-}2.1 \pm~0.3$ & \phantom{0}$-672 \pm \phantom{00}98$ & $\phantom{-}1475 \pm \phantom{0}274$ \nl
Ly~$\alpha$~n  & 1215.67 & $21.3 \pm 3.6$ & \phantom{000}$-7 \pm \phantom{00}19$ & $\phantom{-0}892 \pm \phantom{00}82$ \nl
Ly~$\alpha$~b  & 1215.67 & $36.9 \pm 5.6$ & \phantom{0}$-587 \pm \phantom{0}125$ & $\phantom{-}3086 \pm \phantom{0}232$ \nl
N~V            & 1240.15 & $16.0 \pm~0.5$ & \phantom{0}$-321 \pm \phantom{00}45$ & $\phantom{-}2646 \pm \phantom{0}103$ \nl
C~II           & 1334.53 & $\phantom{-}1.9 \pm~0.3$ & \phantom{0}$-195 \pm \phantom{00}89$ & $\phantom{-}1244 \pm \phantom{0}161$ \nl
Si~IV+O~IV]    & 1400.00 & $\phantom{-}4.7 \pm~0.4$ & $\phantom{-00}13 \pm \phantom{0}109$ & $\phantom{-}2429 \pm \phantom{0}260$ \nl
N~IV]          & 1486.50 & $\phantom{-}2.5 \pm~0.3$ & \phantom{0}$-853 \pm \phantom{0}133$ & $\phantom{-}1801 \pm \phantom{0}229$ \nl
C~IV~n         & 1549.05 & $15.3 \pm 1.7$ & $\phantom{-000}9 \pm \phantom{00}33$ & $\phantom{-}1233 \pm \phantom{00}83$ \nl
C~IV~b         & 1549.05 & $14.2 \pm 1.7$ & \phantom{0}$-614 \pm \phantom{0}125$ & $\phantom{-}3086 \pm \phantom{0}232$ \nl
He~II          & 1640.50 & $12.2 \pm~0.6$ & \phantom{0}$-370 \pm \phantom{00}42$ & $\phantom{-}1750 \pm \phantom{0}104$ \nl
O~III]          & 1664.00 & $\phantom{-}<1.7$ & \phantom{0}$-585 \pm \phantom{0}655$ & $\phantom{-}1750 \pm \phantom{0}104$ \nl
N~III]         & 1750.00 & $\phantom{-}2.3 \pm~0.5$ & $\phantom{-0}145 \pm \phantom{0}120$ & $\phantom{-}2212 \pm \phantom{0}246$ \nl
\enddata
\tablenotetext{a}{Velocities are relative to a systemic redshift of 
$z = 0.0038$ (\cite{Huchra92}).}
\end{deluxetable}

\begin{deluxetable}{lrccc}
\tablecolumns{5}
\tablewidth{0pc}
\tablecaption{Emission Lines in HUT Observation B of \ngc
\label{tbl:B} }
\tablehead{
        \colhead{} & 
        \colhead{$\rm{\lambda_{vac}}$} & 
        \colhead{Flux} & 
        \colhead{$\Delta \upsilon$\tablenotemark{a}} & 
        \colhead{FWHM} \\
        \colhead{Line} & 
        \colhead{$\rm{(\AA)}$} & 
        \colhead{$\rm{(10^{-13}~ergs~cm^{-2}~s^{-1})}$} & 
        \colhead{$\rm{(km~s^{-1})}$} & 
        \colhead{$\rm{(km~s^{-1})}$} }
\startdata
C~III          & \phantom{0}977.03 & $\phantom{-}6.7 \pm~0.4$ & \phantom{0}$-435 \pm \phantom{00}51$ & $\phantom{-}1837 \pm \phantom{0}129$ \nl
N~III          & \phantom{0}991.00 & $\phantom{-}4.4 \pm~0.4$ & \phantom{00}$-77 \pm \phantom{00}81$ & $\phantom{-}2099 \pm \phantom{0}213$ \nl
Ly$\beta$      & 1025.72 & $\phantom{-}5.3 \pm~0.4$ & \phantom{0}$-456 \pm \phantom{00}49$ & $\phantom{-}1750 \pm \phantom{00}87$ \nl
O~VI           & 1031.93 & $10.8 \pm~0.4$ & \phantom{0}$-455 \pm \phantom{00}49$ & $\phantom{-}1750 \pm \phantom{00}87$ \nl
O~VI           & 1037.62 & $12.6 \pm~0.5$ & \phantom{0}$-455 \pm \phantom{00}49$ & $\phantom{-}1750 \pm \phantom{00}87$ \nl
He~II          & 1085.15 & $\phantom{-}3.4 \pm~0.3$ & \phantom{0}$-678 \pm \phantom{00}62$ & $\phantom{-}1557 \pm \phantom{0}131$ \nl
Ly~$\alpha$~n  & 1215.67 & $45.6 \pm 2.8$ & \phantom{00}$-86 \pm \phantom{00}11$ & $\phantom{-0}977 \pm \phantom{00}37$ \nl
Ly~$\alpha$~b  & 1215.67 & $37.8 \pm 4.4$ & \phantom{0}$-663 \pm \phantom{00}61$ & $\phantom{-}3041 \pm \phantom{0}120$ \nl
N~V            & 1240.15 & $23.9 \pm~0.5$ & \phantom{0}$-511 \pm \phantom{00}26$ & $\phantom{-}2588 \pm \phantom{00}63$ \nl
C~II           & 1334.53 & $\phantom{-}2.6 \pm~0.2$ & \phantom{0}$-186 \pm \phantom{00}48$ & $\phantom{-}1089 \pm \phantom{0}120$ \nl
Si~IV+O~IV]    & 1400.00 & $\phantom{-}7.9 \pm~0.4$ & \phantom{0}$-341 \pm \phantom{00}84$ & $\phantom{-}3131 \pm \phantom{0}165$ \nl
N~IV]          & 1486.50 & $\phantom{-}3.3 \pm~0.3$ & \phantom{0}$-928 \pm \phantom{00}87$ & $\phantom{-}1767 \pm \phantom{0}199$ \nl
C~IV~n         & 1549.05 & $24.5 \pm 1.4$ & \phantom{0}$-159 \pm \phantom{00}16$ & $\phantom{-}1127 \pm \phantom{00}47$ \nl
C~IV~b         & 1549.05 & $22.3 \pm 1.5$ & \phantom{0}$-736 \pm \phantom{00}61$ & $\phantom{-}3041 \pm \phantom{0}120$ \nl
He~II          & 1640.50 & $17.4 \pm~0.6$ & \phantom{0}$-576 \pm \phantom{00}31$ & $\phantom{-}1901 \pm \phantom{00}63$ \nl
O~III]          & 1664.00 & $\phantom{-}<1.7$ & \phantom{0}$-189 \pm \phantom{0}350$ & $\phantom{-}1901 \pm \phantom{00}63$ \nl
N~III]         & 1750.00 & $\phantom{-}2.7 \pm~0.4$ & \phantom{00}$-78 \pm \phantom{00}81$ & $\phantom{-}2099 \pm \phantom{0}213$ \nl
\enddata
\tablenotetext{a}{Velocities are relative to a systemic redshift of 
$z = 0.0038$ (\cite{Huchra92}).}
\end{deluxetable}

\section{Intercomparison of the HUT Observations}

Aperture location B, centered on the ionization cone, shows both brighter far-UV
line and continuum emission than either of the other two aperture locations.  
This is readily seen in Figure~3,
which compares the short wavelength portions of the HUT spectra.
To make a more quantitative comparison, we calculated
emission line and continuum flux ratios using the data
from location B as a fiducial reference.
Figures 4 and 5
show the relative emission line and continuum fluxes, respectively.

We can use the relative intensities seen at the various aperture locations
to infer the location and extent of the far-UV line and continuum emission
in NGC~1068.
For this analysis we also use the Astro-1 data from
Kriss et al. (1992).\markcite{Kriss92}
Since these data were obtained through even larger apertures ($18''$ and $30''$
diameters), they provide an overall normalization for the total flux.
(We assume that the fluxes have not varied since our 1990 Astro-1 observations.)
Qualitatively, it is helpful to first consider how one would
expect the ratios to be behave in some simple, limiting cases.
Referring to the relative aperture locations illustrated in
Figure \ref{n1068f1.ps}, one can see that a point source at the location
of the optical nucleus would show roughly the same intensity at all three
aperture locations.
If one offset a point source to the northeast, toward the center of aperture B,
its intensity in aperture B would brighten slightly as vignetting due to
the aperture edge decreased.  Conversely, its intensity as seen through
aperture locations A1 and A2 would decrease drastically as the source moved
beyond the edge of the aperture.
Comparisons to the observed Astro-1 intensities mainly constrain the extent of
the emission source.  In the limit of a large, uniformly
bright extended source, the intensities seen through apertures
A1, A2, and B would all be comparable, but they would
be fainter than that seen in the Astro-1 observation by the ratio of the
angular areas, $( 12'' / 18'')^2 = 0.44$.
For sources with angular extents smaller than the aperture sizes, the ratio
relative to Astro-1 would gradually approach unity, depending on how close
the center of emission was to the aperture edge.

As one can see in Figure~4 and Table~6,
the ratios of the line fluxes among the observations is remarkably uniform.
We find error-weighted averages for the emission line flux ratios
to be $\rm A1/B=0.42\pm0.03$ and \rm $A2/B=0.67\pm0.02$.
Within this range none of the measured flux ratios
differ appreciably (Fig.~4).  
The far-UV line emission distribution appears largely
independent of species or degree of ionization.
In particular, high-excitation-temperature lines like 
\ion{C}{3}~$\lambda 977$, \ion{N}{3}~$\lambda 991$, 
and \ion{O}{6}~$\lambda\lambda 1034,1037$ appear to have
comparable spatial distributions, and their relative brightness seen through
aperture B suggests that most of this emission originates from a region
northeast of the optical nucleus, closer to the center of aperture B.

\vbox to 3.8in {
\vbox to 14pt { \vfill }
\plotfiddle{"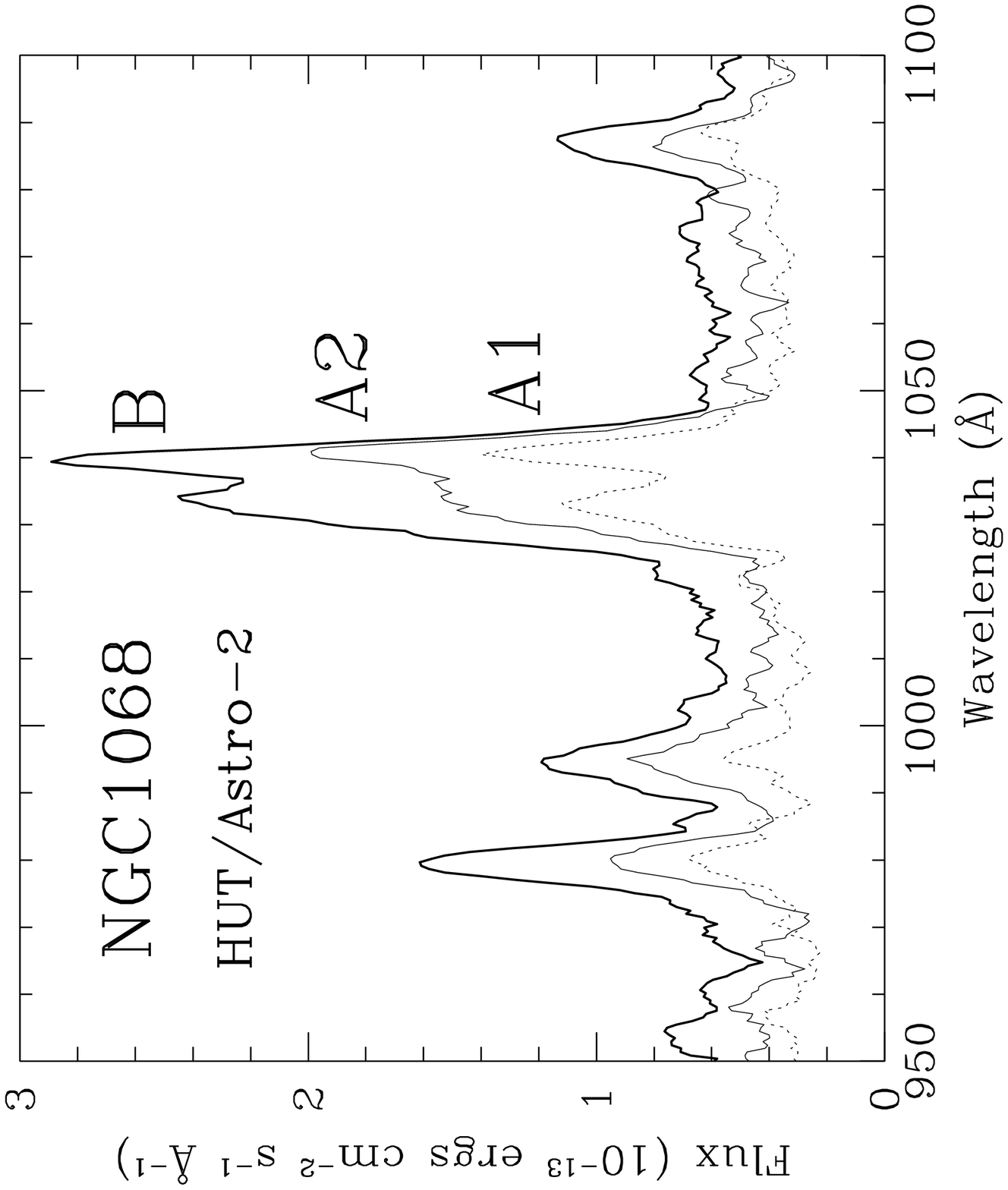"}{2.7 in} {-90}{43}{43}{-195}{240}
\parbox{3.5in}{
\small\baselineskip 9pt
\footnotesize
\indent
{\sc Fig.}~3.---
Comparison of the three HUT spectra in the 950--1100 \AA\ range.
The spectra have been smoothed with a box-car filter of width 5 pixels,
but they have have not been renormalized or scaled.  
The high intensity of the B spectrum, where the aperture was centered on
the ionization cone, relative to A1 and A2, where the ionization cone was
largely outside the aperture, is apparent. 
Note also the greater relative brightness of the emission lines in aperture B
compared to the continuum.
}
}
\setcounter{figure}{3}

\begin{deluxetable}{lrrrr}
\tablecolumns{5}
\tablewidth{0pc}
\tablecaption{HUT Emission Line Intensity Ratios
\label{tbl:emissionratios} }
\tablehead{
	\colhead{Line} &
        \colhead{$\lambda_{vac}$} & 
	\colhead{A1/B} &
	\colhead{A2/B} &
	\colhead{B/Astro-1} \\
	\colhead{} & \colhead{(\AA)} & \colhead{} & \colhead{} & \colhead{}\\ }
\startdata
\ion{C}{3}        & 977.03  & 0.28$\pm$0.05 & 0.47$\pm$0.06& 0.88$\pm$0.17 \\
\ion{N}{3}        & 991.00  & 0.20$\pm$0.05 & 0.73$\pm$0.10& 1.13$\pm$0.31 \\
Ly$\beta$         & 1025.72 & 0.54$\pm$0.12 & 0.66$\pm$0.09& 0.43$\pm$0.09 \\
\ion{O}{6}        & 1031.93 & 0.29$\pm$0.03 & 0.64$\pm$0.05& 0.66$\pm$0.09 \\
\ion{O}{6}        & 1037.62 & 0.41$\pm$0.04 & 0.68$\pm$0.05& 0.60$\pm$0.07 \\
\ion{He}{2}       & 1085.15 & 0.53$\pm$0.09 & 0.63$\pm$0.11& 0.92$\pm$0.26 \\
\Ly narrow        & 1215.67 & 0.25$\pm$0.05 & 0.47$\pm$0.10& 0.45$\pm$0.03 \\
\Ly broad         & 1215.67 & 0.29$\pm$0.10 & 0.98$\pm$0.10& 1.59$\pm$0.35 \\
\ion{N}{5}        & 1240.15 & 0.43$\pm$0.02 & 0.67$\pm$0.03& 0.85$\pm$0.05 \\
\ion{C}{2}        & 1334.53 & 0.26$\pm$0.10 & 0.71$\pm$0.13& 0.74$\pm$0.18 \\
\ion{Si}{4}+\ion{O}{4}] & 1400$\phantom{.50}$ & 0.32$\pm$0.05 & 0.59$\pm$0.06& 0.90$\pm$0.14 \\
\ion{N}{4}]       & 1486.50 & 0.61$\pm$0.10 & 0.76$\pm$0.13& 0.65$\pm$0.15 \\
\ion{C}{4} narrow & 1549.05 & 0.41$\pm$0.08 & 0.63$\pm$0.06& 0.62$\pm$0.06 \\
\ion{C}{4} broad  & 1549.05 & 0.38$\pm$0.08 & 0.64$\pm$0.07& 1.23$\pm$0.25 \\
\ion{He}{2}       & 1640.50 & 0.46$\pm$0.03 & 0.70$\pm$0.04& 0.81$\pm$0.07 \\
\ion{N}{3}]       & 1750.00 & 0.65$\pm$0.17 & 0.84$\pm$0.24& 0.47$\pm$0.14 \\
\enddata
\end{deluxetable}

The intensity ratios of the narrow to broad components for the
\Ly and \ion{C}{4}~$\lambda 1549$ emission lines 
show little variance among the three observations or between the \Ly and
\ion{C}{4} lines themselves (Table~7).
A weighted average of the three pointings and both the \Ly and \ion{C}{4} 
emission lines yields $I_{narrow}/ I_{broad}=1.09\pm0.06$.
In contrast, previous observations by Kriss et al. (\markcite{Kriss92}1992)
through an 18\arcsec~aperture
found a more dominant contribution from the narrow lines:
$I_{\rm{Ly}\alpha,narrow}/I_{\rm{Ly}\alpha,broad}=4.28\pm0.19$
and $I_{\rm{CIV},narrow}/I_{\rm{CIV},broad}=2.19\pm0.20$.
This is likely due to the smaller aperture used in the Astro-2 observations.
These new observations include less of the extended narrow line region.
Also, in Kriss et al. (\markcite{Kriss92}1992) the \Ly
and \ion{C}{4} narrow-to-broad line ratios 
are very different, not uniform as in our observations.
A possible explanation could be that the \ion{C}{4}
narrow-line region is less extended than the
\Ly narrow-line region.

\vbox to 3.5in {
\plotfiddle{"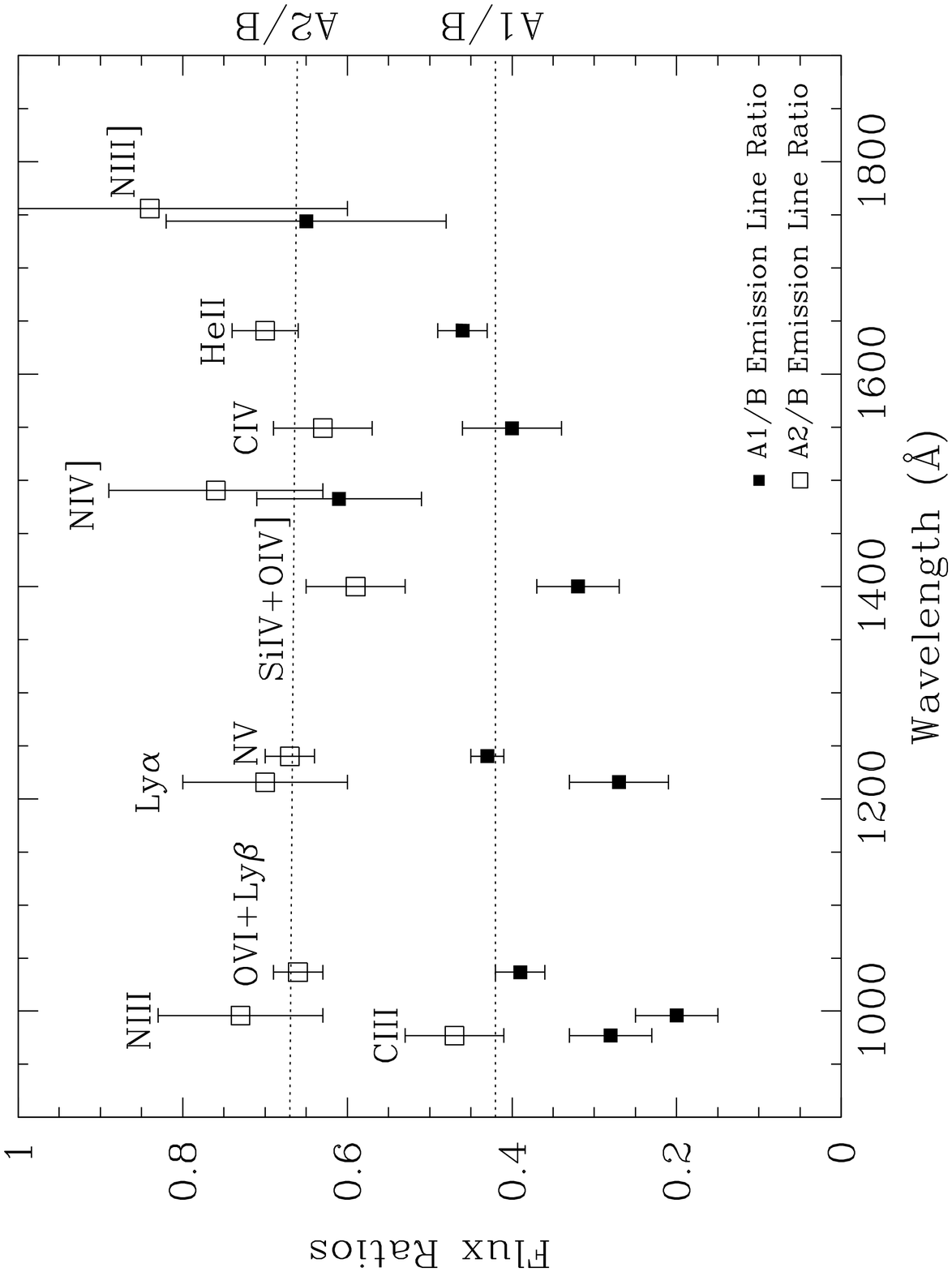"}{2.6 in} {-90}{33}{33}{-125}{200}
\parbox{3.5in}{
\small\baselineskip 9pt
\footnotesize
\indent
{\sc Fig.}~4.---
Flux ratios of selected HUT emission lines
are shown for aperture location A1 relative to B (A1/B) and 
A2 relative to B (A2/B).
The dotted lines are the weighted mean ratios for A1/B and for A2/B.
}
}
\setcounter{figure}{4}

\vbox to 3.3in {
\plotfiddle{"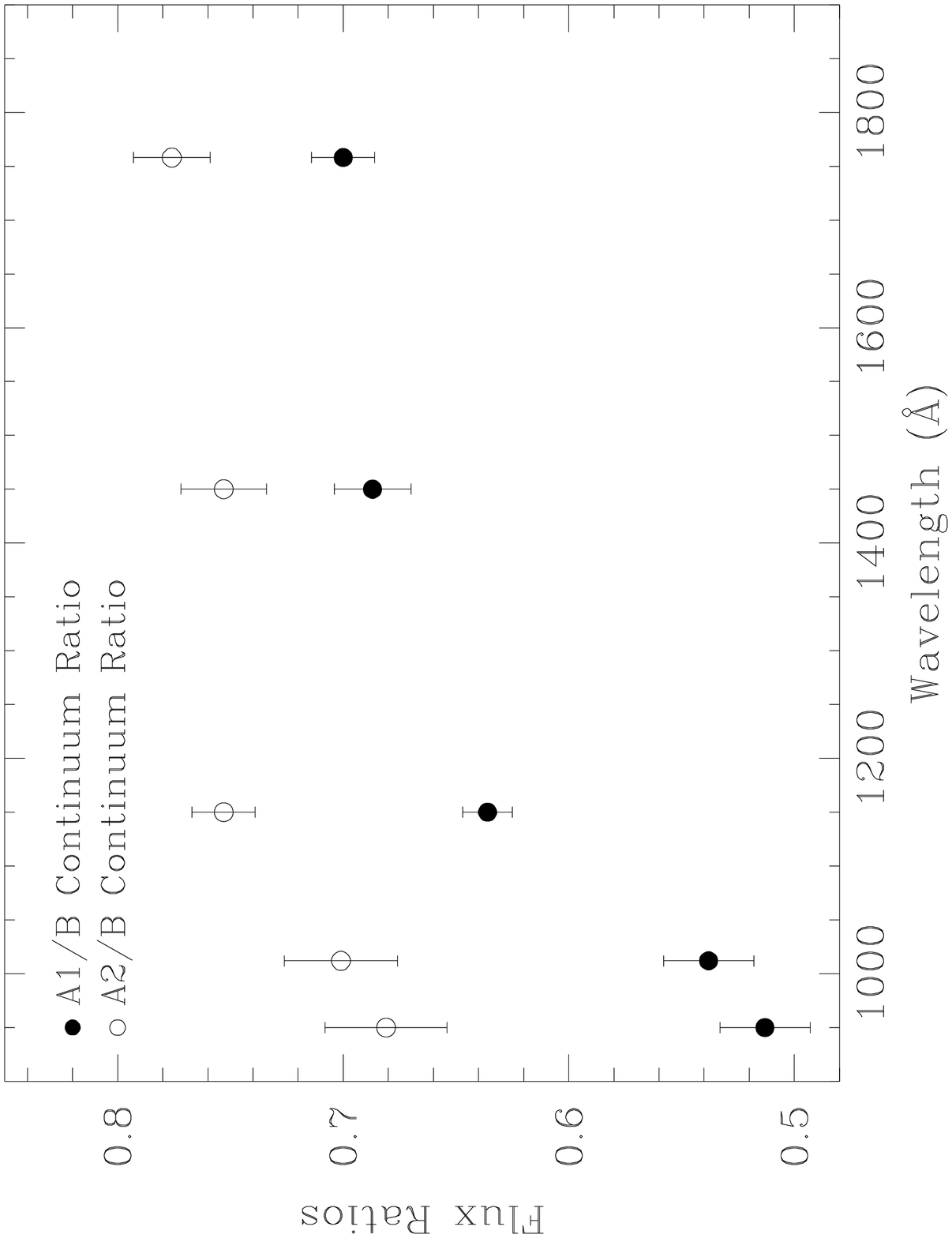"}{2.7 in} {-90}{35}{35}{-130}{215}
\parbox{3.5in}{
\small\baselineskip 9pt
\footnotesize
\indent
{\sc Fig.}~5.---
HUT continuum flux ratios are shown for aperture locations
A1 and A2 relative to B.
}
}
\setcounter{figure}{5}

\begin{center}
\small
{\sc Table 7\\
Intensity Ratios of Narrow to Broad Lines}
\vskip 4pt
\begin{tabular}{lrrr}
\hline
\hline
	 & \multicolumn{3}{c}{Observation}  \\
   Line  & A1  & A2  & B  \\
\hline
\Ly  & $1.05\pm0.38$	& $0.58\pm0.13$ & $1.21\pm0.16$ \\
C~IV & $1.18\pm0.33$    & $1.08\pm0.17$	& $1.11\pm0.10$ \\
\hline
\end{tabular}
\vskip 4pt
\end{center}
\setcounter{table}{7}

\begin{deluxetable}{ccrr}
\tablecolumns{4}
\tablewidth{0pc}
\tablecaption{HUT Continuum Intensity Ratios
\label{tbl:HUTcontinuumratios} }
\tablehead{
	\colhead{Wavelength Region} &
        \colhead{Number of} & 
	\colhead{A1/B} &
	\colhead{A2/B} \\
	\colhead{\AA} & \colhead{Data Points} & \colhead{} & \colhead{}\\ }
\startdata
$\phantom{0}$925--$\phantom{0}$970	& \phantom{0}89 & 0.513$\pm$0.020 & 0.681$\pm$0.027 \\
1004--1020	        & \phantom{0}40 & 0.538$\pm$0.020 & 0.701$\pm$0.025 \\ 
1100--1200        	&       194 & 0.636$\pm$0.011 & 0.753$\pm$0.014 \\
1422--1469        	& \phantom{0}92 & 0.687$\pm$0.017 & 0.753$\pm$0.019 \\
1674--1842        	&       327 & 0.700$\pm$0.014 & 0.776$\pm$0.017 \\
\enddata
\end{deluxetable}

\begin{deluxetable}{lcccc}
\tablecolumns{5}
\tablewidth{0pc}
\tablecaption{UV Emission Region Model Parameters
\label{tbl:hutloc} }
\tablehead{
	\colhead{Wavelength Region} &
        \colhead{$\Delta \alpha$\tablenotemark{a}} & 
	\colhead{$\Delta \delta$\tablenotemark{b}} &
	\colhead{FWHM} &
	\colhead{$\chi^2$} \\
	\colhead{} &
	\colhead{(arc sec)} & \colhead{(arc sec)} & \colhead{(arc sec)} &
	\colhead{} \\ }
\startdata
Emission lines	& $+2.04\pm0.05$ & $-0.27\pm0.10$ & 5.5$\pm$0.7 & 0.12 \\
\phantom{0}925--\phantom{0}970 \AA & $+1.31\pm0.15$ & $+0.09\pm0.10$ & 5.5$\pm$0.6 & 0.55 \\
1004--1020 \AA	& $+1.27\pm0.23$ & $+0.09\pm0.23$ & 6.0$\pm$0.4 & 0.08 \\
1100--1200 \AA	& $+0.86\pm0.10$ & $+0.23\pm0.10$ & 7.5$\pm$0.5 & 0.08 \\
1422--1469 \AA	& $+0.59\pm0.15$ & $+0.50\pm0.15$ & 9.0$\pm$0.5 & 0.04 \\
1674--1842 \AA	& $+0.41\pm0.10$ & $+0.41\pm0.10$ & 7.5$\pm$0.4 & 0.22 \\
\enddata
\tablenotetext{a}{Offset in Right Ascension relative to NGC 1068's optical
nucleus.}
\tablenotetext{b}{Offset in Declination relative to NGC 1068's optical nucleus.}
\end{deluxetable}

In contrast to the emission lines,
Figure~5 and Table~8
show that the far-UV continuum has a very different 
distribution from that of the far-UV emission lines.
The continuum is concentrated in location B,
but not nearly as strongly as the emission lines.
Also, unlike the emission lines, the continuum ratios {\it are} wavelength
dependent.
The continuum appears to have a broader spatial distribution
covering all three aperture locations at wavelengths longward of 1200 \AA.
The continuum flux becomes increasingly
concentrated in the ionization cone region at wavelengths shortward of 1200 \AA.

To quantify these inferences, we fit a simple model of the line and
continuum flux surface brightness distributions to the relative intensities
as seen at the different aperture locations.
The total flux from Astro-1 and the three Astro-2 measurements provide us with
four data points.
If we model the emission with a Gaussian surface brightness distribution,
we have four free parameters that are exactly constrained by our data---
total intensity, location (two coordinates), and the full-width at half maximum
(FWHM).
We determine the parameters by using a $\chi^2$ fit to a series of Gaussian
images with the FWHM incremented in steps of $0.5''$.
At each step we vary the location freely and measure the total fluxes
inside circular apertures $12''$ in diameter whose centers were fixed at the
relative locations determined by the HUT pointing errors shown in
Figure~\ref{n1068f1.ps}.

Table~9 gives the best-fit locations relative to the optical
nucleus and the sizes for the emission line and continuum flux emitting regions.
The near-zero values of $\chi^2$ in our fits result from the zero degrees of
freedom--- four data points precisely determine four free parameters.
Using $\chi^2$ statistics gives the added value of permitting us to determine
error bars for our results.
The quoted error bars are $1 \sigma$ (assuming $\Delta\chi^2 = 1$ relative
to $\chi^2_{min}$).  These errors are appropriate for {\it internal}
comparisons of relative locations and sizes; however, they do not include
the systematic errors of $\sim 0.5''$ in determining the location of the optical
nucleus in the HUT acquisition video frames.

The results of the fits roughly correspond to our qualitative conclusions.
Since the point-spread function (PSF) of HUT is $\sim4''$ (\cite{Davidsen92}),
we see that the emission line region, with FWHM=$5.5''$, is slightly resolved.
In contrast, the far-UV continuum comes from an extended region spanning many
arc seconds.  At wavelengths longer than 1200 \AA, the center of the emission
is consistent with that of the optical nucleus.  At shorter wavelengths,
the peak of the emission shifts in the direction of the emission line region,
and its size becomes comparable to that of the emission line region.

\begin{figure*}[t]
\plotfiddle{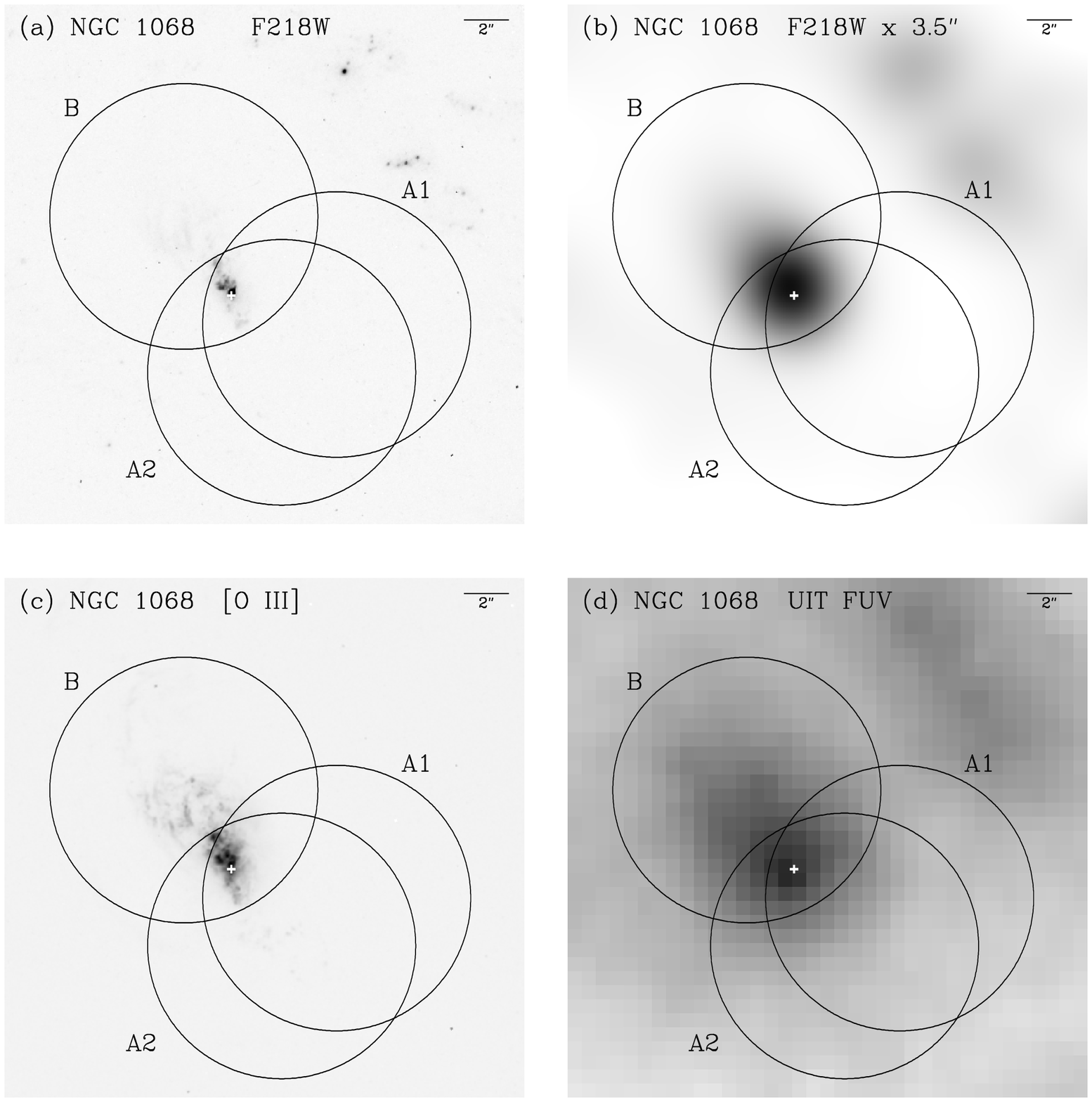}{6.7 in} {0}{85}{85}{-260}{-76}
\caption{
Derived locations of the three HUT apertures superimposed on HST/WFPC2 and
UIT FUV images.
Panel (a) shows the full-resolution HST/WFPC2 F218W image;
panel (b) is the same image convolved to the resolution of HUT
with a Gaussian of FWHM=$3.5''$;
panel (c) shows the full-resolution HST/WFPC2 [O~III] $\lambda 5007$ image;
panel (d) shows the stacked high-resolution UIT FUV image.
North is up, and East is to the left.
The crosses mark the location of the optical nucleus.
\label{n1068f6.ps}}
\end{figure*}

The relative offsets of the different regions are a bit more puzzling.
We see that the center of the emission line region lies $\sim1.5''$ east
of the long wavelength UV continuum, in the direction of the center of aperture
B and the ionization cone.  The puzzling aspect is the $\sim 0.7''$ offset
to the south.  Given the geometry of the ionization cone, we would have
expected a slight shift to the north.
The puzzle is most likely explained by a more detailed consideration of the
actual morphology.
As one can see in Fig.~6, the northern edge of aperture A1
lies near the starburst knots visible in the HST and UIT images.
Some flux from these knots falls in the aperture due to the broad PSF and
pointing jitter.  This contributes to the large size inferred for the continuum
regions at wavelengths longward of 1200 \AA, and also shifts the fitted
centroid to the north.
This conclusion is made clearer through a detailed comparison to the HST images
that we discuss in the next section.

\section{Comparison to HST and UIT Images}

To reference the coarse spatial resolution of our far-UV 1-D aperture spectra
to higher resolution, longer wavelength images, we use archival
HST WFPC2 images of \ngc obtained by H. Ford, as reduced and analyzed by
Dressel et al. (1997)\markcite{Dressel97}.
We also compare our results to the far-UV image obtained by
Neff et al. (1994\markcite{Neff94}) using UIT on the Astro-1 mission.
Table~10 summarizes nine images covering a range
of emission lines and continuum bands from the far-UV to the visible.

To compare the HUT spectra to the HST and UIT images,
the images must be registered on a common coordinate system,
convolved to the HUT spatial resolution,
and integrated over the spectrograph entrance aperture.
We start with the F218W ultraviolet continuum image as this overlaps the
long wavelength end of the HUT bandpass and is devoid of strong emission lines.
As noted in the last section, the HUT PSF has FWHM$\sim4''$.
To obtain a more precise estimate specifically for the NGC~1068 observations
that also takes into account the pointing jitter,
we convolved the F218W image with a series of Gaussians,
incrementing the FWHM in steps of $0.5''$.
As in our fits in the last section,
we measured total fluxes inside circular apertures $12''$ in diameter
corresponding to the HUT aperture locations.
By comparing the intensity ratios at the three locations measured in the 
convolved HST image to those in the HUT spectra in the 1674--1842 \AA\ 
continuum region, we are able to determine the best fit registration
for the HUT apertures on the HST images as well as the best matching PSF.
The best-fit Gaussian has FWHM=$3.5'' \pm 0.5''$.
The registration places the peak of the UV emission in the F218W image at a
position relative to the optical nucleus as determined from the HUT video frames
of $\Delta\alpha = +0.23'' \pm 0.14''$, $\Delta\delta = -0.36'' \pm 0.14''$.
Note that this reverses the situation encountered in the last section, where
the fit to the HUT data alone placed the centroid of the 1800 \AA\ continuum
flux {\it north} of the optical nucleus.
Figure~6 shows the HUT aperture locations determined in this
process superimposed on the F218W image of \ngc.  
By examining Fig.~6b, where the HST image is convolved with the
HUT PSF, one can see that the starburst knots northwest of the nuclear region
contribute a significant amount of flux to aperture A1.
This is the likely explanation for the width inferred from the simple
Gaussian fit in the last section as well as the bias to the north for the
centroid of that fit.

\begin{center}
\small
{\sc Table 10\\
UIT and HST/WFPC2$\rm ^a$ Images of NGC~1068}
\vskip 4pt
\begin{tabular}{llc}
\hline
\hline
Name & Notes & Exposure Time \\
      &       & ($s$)\\
\hline
UIT/B1      & 1250--2000 \AA\              & 1629 \\
WFPC2/F218W & continuum                    & 1200 \\
WFPC2/F336W & continuum                    & 450 \\
WFPC2/F343N & [\ion{Ne}{5}] $\lambda 3425$ & 900 \\ 
WFPC2/F375N & [\ion{O}{2}]  $\lambda 3727$ & 900 \\
WFPC2/F502N & [\ion{O}{3}]  $\lambda 5007$ & 450 \\ 
WFPC2/F547M & continuum                    & 220 \\
WFPC2/F656N & \ha                          & 450 \\
WFPC2/F791W & continuum                    & 220 \\
\hline
\end{tabular}
\vskip 2pt
\parbox{3.5in}{
\footnotesize
$\rm ^a$HST/WFPC2 images courtesy of Zlatan Tsvetanov.
}
\end{center}

The flux distribution of the F218W image also provides a good fit to the
1450 \AA\ region flux in the HUT data.  With a FWHM of $5.0'' \pm 0.5''$,
we obtain $\chi^2 = 0.07$.
The F218W spatial distribution, however, gives a significantly worse fit to the
short wavelength fluxes in the 1004--1020 \AA\ band.
Here we obtain $\chi^2 = 2.80$ for a best-fit FWHM of $2.0'' \pm 0.5''$.
This is consistent with the results of our simpler, single Gaussian model
that showed the short-wavelength radiation arising in a more compact
region--- the flux distribution in the F218W image has a greater spatial
extent than is seen through the HUT apertures at 1000 \AA.
The F218W spatial distribution is a very poor match to the emission-line
flux distribution--- $\chi^2 = 113.36$ for the best fit.
This rules out any substantial contribution of the northwest starburst knots
to the far-UV line emission seen with HUT.

If we do a fit to optimize the registration and resolution of the 
[\ion{O}{3}] image,
we obtain $\chi^2 = 0.03$ for FWHM=$5.0 \pm 0.4$ arc sec and an
offset relative to the position of the F218W image of
$\Delta\alpha = +1.40'' \pm 0.10''$, $\Delta\delta = 0.59'' \pm 0.10''$.
Panel (c) of Figure~\ref{n1068f6.ps} shows the HUT aperture locations
superimposed on the [\ion{O}{3}] image of \ngc.  
These results imply the UV line emission has a broader spatial distribution
than that of the [\ion{O}{3}] $\lambda 5007$ emission, and it is offset
to the northeast.

The UIT image is described by Neff et al. (1994\markcite{Neff94}).
We obtained the six $2048 \times 2048$ images made using the far-UV detector
through the B1 Sr$\rm F_2$ filter from the
National Space Sciences Data Center (NSSDC).
These images had exposure times ranging from 22 s to 752 s, with the shortest
images providing unsaturated exposures of the brightest portions of the
nuclear region.
We registered and stacked these images to produce the image shown in the
lower right, Fig.~6d.
Comparing the UIT image to the convolved HST F218W image, one sees that
while similar, the UIT image has distinctly brighter emission to the NE of
the nucleus in the ionization-cone region.
This is likely due to the fact that the UIT image includes bright line emission
in its bandpass from Si~{\sc iv}$\lambda 1400$, C~{\sc iv}$\lambda 1549$,
He~{\sc ii}$\lambda 1640$, and C~{\sc iii}]$\lambda 1909$.
The enhanced emission to the NE resembles the cone defined by the [{\sc O~iii}]
$\lambda 5007$ emission shown in Fig.~6c.

Finally, we convolved the other seven HST images to the HUT spatial resolution
using a Gaussian with FWHM=$3.5''$
and measured relative intensities by integrating over
the three HUT aperture locations using the registration inferred from our fit
to the F218W image.
The resulting ratios are shown in Figure~7
and listed in Table~11.
The continuum flux ratios computed from the HST images continue
the trend with wavelength apparent in the HUT data.
The observed ratios are consistent with a broader, less concentrated source
of radiation becoming more dominant at longer wavelengths.
The distribution of line emission viewed with HST is qualitatively, but
not precisely, similar to what we see with HUT.
The very broad spatial distribution of the low ionization emission lines,
H$\alpha$ and [\ion{O}{2}] $\lambda 3727$, readily apparent in the HST images,
also shows up in the ratios illustrated in Figure~7
and Table~11.
The closest resemblance is for the highest ionization lines,
[\ion{O}{3}] $\lambda 5007$ and [\ion{Ne}{5}] $\lambda 3425$, but even
here we do not get a good match to the flux distribution measured with HUT.

\vbox to 3.1in {
\vbox to 14pt{ \vfill }
\plotfiddle{"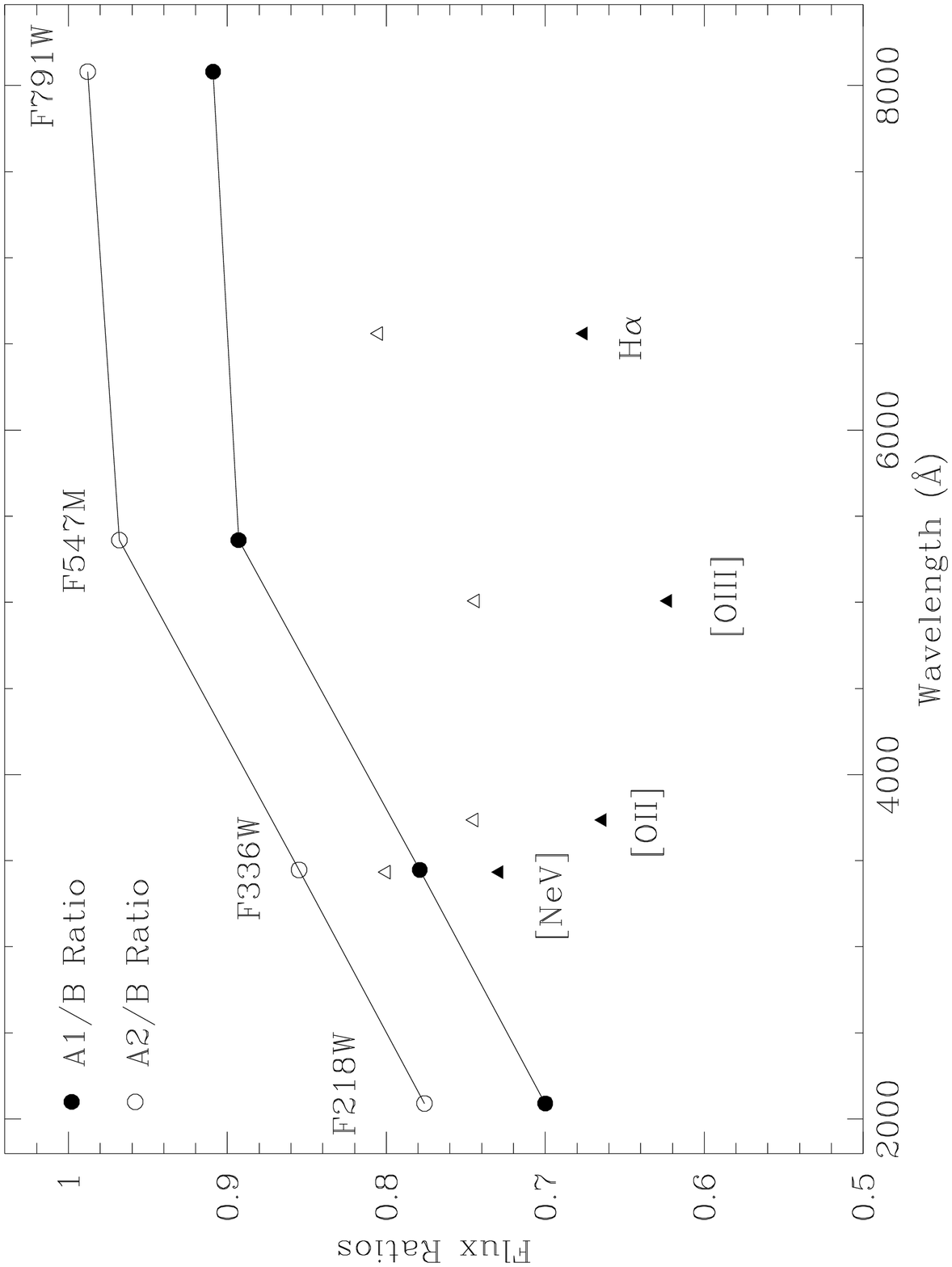"}{2.5 in} {-90}{35}{35}{-133}{210}
\parbox{3.5in}{
\small\baselineskip 9pt
\footnotesize
\indent
{\sc Fig.}~7.---
Emission line and continuum flux ratios
derived from HST images convolved to the 
HUT resolution with a Gaussian of 3.5$''$ FWHM.  Fluxes are measured
within 12$''$ circular apertures at the HUT observation
locations determined from the fit to the F218W image.
}
}
\setcounter{figure}{7}

\section{Discussion}

\subsection{Line Emission and Potential Excitation Mechanisms}

The greater extent and larger offset to the northeast of the far-UV line
emission relative to the [\ion{O}{3}] $\lambda 5007$ emission
are at odds with a photoionized origin for the far-UV emission lines.
As argued by Kriss et al. (1992)\markcite{Kriss92}, the strengths of the
temperature-sensitive \ion{C}{3} $\lambda 977$ and \ion{N}{3} $\lambda 991$
emission lines imply high temperatures for the line-emitting gas.
For photoionized excitation, this implies high ionization parameters.
Hence, one would expect the \ion{C}{3} $\lambda 977$ and
\ion{N}{3} $\lambda 991$ emission to come from a more compact region closer to
the central ionizing source than the [\ion{O}{3}] $\lambda 5007$ emission.
Instead, the offset to the northeast along the direction of the radio jet
suggests that interaction of the jet with the line emitting clouds
has more importance for the production of these emission lines than
photoionization by the central engine.

\begin{center}
\small
{\sc Table 11\\
HST Line and Continuum Intensity Ratios}
\vskip 4pt
\begin{tabular}{lccc}
\hline
\hline
Line/Filter       & Wavelength  & A1/B   &  A2/B \\
\hline
{[\ion{Ne}{5}]}   & 3430	& 0.729  &  0.800 \\
{[\ion{O}{2}]}    & 3736	& 0.664  &  0.745 \\
{[\ion{O}{3}]}    & 5012	& 0.623  &  0.744 \\ 
H$\alpha$         & 6562	& 0.676  &  0.805 \\
F218W             & 2189        & 0.700  &  0.776 \\
F336W             & 3342        & 0.779  &  0.855 \\
F547M             & 5476        & 0.893  &  0.968 \\
F791W             & 7926        & 0.909  &  0.988 \\
\hline
\end{tabular}
\end{center}

The fluxes measured in the \ion{C}{3} $\lambda 977$ and \ion{N}{3} $\lambda 991$
emission lines in our Astro-2 observations are consistent with those seen in
Astro-1, given the geometrical uncertainties.
These uncertainties, the shorter observation times, and the complicating
effects of airglow in the Astro-2 observations actually make the Astro-1
measurements a more reliable measure of the total flux contributing to the
temperature-sensitive ratios I($\lambda 1909$)/I($\lambda 977$) and
I($\lambda 1750$)/I($\lambda 991$).
Using the values with 90\% confidence errors as given by
Kriss et al. (1992)\markcite{Kriss92},
improved atomic physics calculations since then allow us to update the
inferred temperature for the line-emitting gas.
For I($\lambda 1909$)/I($\lambda 977$) = $3.15 \pm 0.51$,
the diagnostic diagrams in McKenna et al. (1999)\markcite{McKenna99} give
$T = 21,700^{+1100}_{-1000}~\rm K$.
For I($\lambda 1750$)/I($\lambda 991$) = $1.46 \pm 0.34$, we obtain
$T = 28,700^{+3800}_{-2400}~\rm K$.
The revisions to the relevant atomic parameters have moved both estimates
in opposite directions--- the \ion{C}{3} temperature is now lower, while
the \ion{N}{3} value is higher.

As noted by Kriss et al. (1992)\markcite{Kriss92}, both temperature estimates
are lower limits since they rely on {\it observed} values and do not include
any corrections for extinction, which may be substantial.
Our fits to the continuum give a low extinction, $\rm E( B - V) = 0.02$,
but, as we argue below, this is not very reliable due to the wide variety
of actual sources for the continuum light and its subsequent shape.
The He~{\sc ii} recombination lines ($\lambda 4686$, $\lambda 1640$, and
$\lambda 1085$) give $\rm E( B - V) = 0.1 - 0.15$ (\cite{Kriss92};
\cite{Koski78}), and other indicators
suggest the extinction to the line emitting regions may be as high as
$\rm E( B - V) = 0.4$ (\cite{Malkan83}).
At values this high, the inferred \ion{C}{3}
temperature is 59,300~K, and the \ion{N}{3} temperature is $> 100,000$~K.

The high temperatures implied by the observed \ion{C}{3} and \ion{N}{3} line
ratios are easily matched by the autoionizing shock models computed by
Allen, Dopita, \& Tsvetanov (1998)\markcite{Allen98}.
The diagnostic diagrams they developed to discriminate between shocks
and photoionization show a clear separation between shock and photoionization
models when one compares the ratios [\ion{O}{3}]/H$\beta$ and either
\ion{C}{3}] $\lambda 1909$/\ion{C}{3} $\lambda 977$ or
\ion{C}{3}] $\lambda 1750$/\ion{C}{3} $\lambda 991$.
For either ratio, the temperatures are higher than can be produced in
typical photoionization models.
Neither is a precise match to any of the simple shock models.
Given that our large apertures encompass a wealth of unresolved complex spatial
structure, it is easy to see that any simple model could easily fail.
A clear resolution of this problem will require sub-arcsecond
observations in far-UV lines.
An experiment optimized for far-UV imaging and long-slit spectroscopy
in the 900--1200 \AA\ band could provide the key data for this problem and
many others.

Ferguson et al. (1995)\markcite{FFP95} argue that higher-than-expected
intensities of \ion{C}{3} $\lambda 977$ and \ion{N}{3} $\lambda 991$
can be produced in photoionized gas by fluorescent processes.
To produce the intensities observed in NGC~1068, however, their models
require turbulent velocities of $> 1000~\rm km~s^{-1}$.
While line widths this high are observed in NGC~1068, it is hard to see
how one can avoid fast shocks in clouds with such high internal turbulence.
The fluorescent enhancement of photoionized emission is formally possible,
but it is also certainly an incomplete physical picture of the excitation
and emission process under such extreme hydrodynamic conditions.
One can also get a velocity spread of $1000~\rm km~s^{-1}$ in an accelerating
wind, but one would expect this to occur in a very small spatial region,
essentially a point source with a size on the order of the electron scattering
mirror.  The spatial distribution inferred from our data definitely excludes
a point-like source, and so we rule out this alternative.

Another frequently cited indicator of shock excitation is
the 1.6435 $\mu$m transition of [\ion{Fe}{2}].
Given the close spatial correlation between the morphology
of the radio jet in NGC~1068 and of the [\ion{Fe}{2}] emission,
Blietz et al. (1994)\markcite{Blietz94} concluded that the emission was
produced in gas irradiated by nuclear X-rays, or in shocks excited by
an outflow or jet from the nucleus.
Examining the morphology of the [\ion{Fe}{2}] emission in their Figure 1,
one can see that it peaks near the optical/IR nucleus, but that a significant
fraction of the emission arises from a more extended region $\sim 1''$
to the northeast along the axis of the radio jet.
This is quite similar to the morphology we infer for the far-UV emission
lines from our own observations.

\subsection{The Origin of the Continuum Radiation}

The strong wavelength dependence of the morphology of the continuum emission
suggests that a variety of sources contribute to the continuum light
seen in the HUT spectra.
This supports the conclusions of Neff et al. (1994\markcite{Neff94}) that
``several sources probably contribute to the integrated UV emission of \ngc,"
including dust and electron-scattered nuclear radiation, starlight, and
line and continuum emission from the NLR.
On spatial scales of 30\arcsec\ and at wavelengths longward of 1200 \AA,
starlight is a major contributor.
Heckman et al. (1995)\markcite{Heckman95} discuss the aperture-size
dependence of the nuclear UV flux from \ngc\ and conclude that apertures
that include the starburst ring contain substantial amounts of UV flux
from starlight.
The broader spatial extent we observe for radiation at wavelengths $>1200$ \AA\
(see \S 4) suggests that much of this light has a stellar origin.
This stellar flux is directly visible in the spectra from
the $30''$-aperture Astro-1 observations (\cite{Kriss92}).
As discussed in \S5 and shown in Fig.~6b, portions of the starburst ring
also contribute to the flux seen in the A1 spectrum.

Below 1200 \AA\ the increasing concentration of the continuum emission in
the vicinity of the ionization cone can be attributed to a stronger relative
contribution of scattered nuclear radiation.
This is likely due to two factors that cannot be disentangled at the
spatial resolution of our observations.
First, some of this light must be due to the electron scattering mirror
inferred from the spectropolarimetric observations
(\cite{Antonucci85}; \cite{MGM91}; \cite{Code93}; \cite{Antonucci94}).
Second, the NE dust cloud visible in the imaging polarimetry of
Miller, Goodrich, \& Matthews (1992)\markcite{MGM91}
and in the UV spectropolarimetry of Code et al. (1993)\markcite{Code93}
falls squarely within the HUT aperture location B.
Since the scattering cross section of dust rises rapidly with shorter
wavelengths, the intensity of scattered light from this cloud will form
an increasingly large fraction of the signal in aperture B at wavelengths
shortward of 1200 \AA.
The roughly $5''$ extent we infer for the 1000 \AA\ continuum light is
comparable to the $5''$ distance of the NE cloud from the optical nucleus,
and could account entirely for the size we infer from our observations.

To assess the extent to which our observed fluxes are due to scattered
radiation from the obscured AGN, we compare our results to spectropolarimetric
observations obtained with HST (\cite{Antonucci94}) and with the
Wisconsin Ultraviolet Photo-Polarimeter Experiment (WUPPE) (\cite{Code93}).
At 1800 \AA, the portion of the UV flux attributable to direct reflection of
the central engine by the electron scattering mirror is
$3.3 \times 10^{-14}~\rm ergs~cm^{-2}~s^{-1}~\AA^{-1}$ (\cite{Antonucci94}).
The NE dust cloud contributes
$\sim 1.1 \times 10^{-14}~\rm ergs~cm^{-2}~s^{-1}~\AA^{-1}$ (\cite{Code93}).
The sum accounts for only $\sim$80\% of the flux we observe at 1800 \AA\ at
aperture position B in our observations.
At aperture locations A1 and A2, this total accounts for 89\% and 73\% of the
observed flux, respectively.
As locations A1 and A2 largely exclude the NE dust cloud, the fraction
of the observed UV radiation we see at these locations attributable to
scattered AGN radiation is likely even less.
Given that the A1 and A2 spectra are significantly redder than the B spectrum,
we conclude that diffusely distributed starlight and/or additional scattered
(but reddened) nuclear flux is present within the central $\sim6''$
surrounding the nucleus.

\section{Conclusions}

We have described three spatially distinct HUT observations of NGC~1068.
During observation A1 the optical nucleus was near the eastern edge of the
aperture while in A2 the nucleus was near the northeastern edge.
In the third observation, B, we centered the aperture on the ionization cone
with the optical nucleus near the southwestern edge.

The observed fluxes and emission line ratios are consistent with those
seen in our Astro-1 observations.
The far-UV emission lines are brightest in aperture B in the
vicinity of the ionization cone and the radio jet.
All the far-UV emission lines have similar spatial distributions,
including the high-excitation-temperature lines
\ion{C}{3} $\lambda977$, \ion{N}{3} $\lambda991$,
and \ion{O}{6} $\lambda\lambda1032,1037$.
We found observation B to be brighter than A1 and A2 in far-UV 
emission lines by factors of $\sim2.4$ and $\sim1.5$ respectively.
From comparison to HST images, we find that
the far-UV emission lines have a spatial distribution
relative to the [\ion{O}{3}] $\lambda5007$ emission that is
more extended and offset further to the northeast along the direction of the
radio jet.

Using updated atomic physics (\cite{McKenna99}) to re-evaluate the
temperatures implied by the ratios I($\lambda 1909$)/I($\lambda 977$) and
I($\lambda 1750$)/I($\lambda 991$),
we find a lower limit from the \ion{C}{3} ratio of 21,200~K,
and a lower limit of 27,200~K from the \ion{N}{3} ratio.
Given the high ionization parameter normally required to produce
such high temperatures in photoionized gas, we would have expected 
the spatial distribution inferred from our observations to be more
compact and more concentrated near the nucleus.
Since it appears more extended and offset along the axis of the radio jet,
we conclude that our Astro-2 observations provide more evidence for this
emission to arise in shock-heated rather than photoionized gas.

The continuum appears to have a broader spatial 
distribution than the emission lines,
but it grows progressively more concentrated in the 
ionization cone region at wavelengths shorter than 1200 \AA.
At longer wavelengths an increasing portion of the flux appears to come
from starlight.  Within aperture location A1, this arises in the starburst
knots northwest of the optical nucleus.
Within aperture B, $\sim$80\% of the flux can be attributed to
scattered nuclear radiation from the electron scattering mirror
and from the NE dust cloud.  The remaining flux must come from
more diffusely distributed starlight or scattered (but reddened)
nuclear radiation within the central $\sim6''$.

\acknowledgements
We are grateful to Z. Tsvetanov for providing the reduced HST images
and to D. Ubol for help in producing Fig. 6.
J. G. appreciates helpful conversations with B. Greeley and R. Telfer.
This research was supported in part by NASA contract
NAS 5-27000 to the Johns Hopkins University
and by NASA Long-Term Space Astrophysics grant NAG 5-3255.

\end{document}